\documentstyle[eqsecnum,aps,prd]{revtex}
\begin{document}
\newcommand{\bfx}{{\bf x}}
\newcommand{\bfy}{{\bf y}}
\newcommand{\bfr}{{\bf r}}
\newcommand{\bfk}{{\bf k}}
\newcommand{\bfa}{{\bf a}}
\newcommand{\bkp}{{\bf k'}}
\newcommand{\order}{{\cal O}}
\newcommand{\beq}{\begin{equation}}
\newcommand{\eeq}{\end{equation}}
\newcommand{\beqn}{\begin{eqnarray}}
\newcommand{\eeqn}{\end{eqnarray}}
\newcommand{\lmk}{\left(}
\newcommand{\rmk}{\right)}
\newcommand{\lkk}{\left[}
\newcommand{\rkk}{\right]}
\newcommand{\lnk}{\left\{}
\newcommand{\rnk}{\right\}}
\newcommand{\call}{{\cal L}}
\newcommand{\calh}{{\cal H}}
\newcommand{\ppp}{\partial}
\newcommand{\tilq}{{\tilde q}}
\newcommand{\tilp}{{\tilde p}}
\newcommand{\tilu}{{\tilde U}}
\newcommand{\tilh}{{\tilde H}}
\newcommand{\baret}{{\bar \eta}}
\newcommand{\xbar}{{\overline x}}
\thispagestyle{empty}
\thispagestyle{empty}
{\baselineskip0pt
\leftline{\large\baselineskip16pt\sl\vbox to0pt{\hbox{\bf DAMTP} 
               \hbox{\bf University of Cambridge}\vss}}
\rightline{\large\baselineskip16pt\rm\vbox to20pt{
               \hbox{DAMTP-153}
               \hbox{UTAP-314}
               \hbox{RESCEU-58/98}           
               \hbox{OCHA-PP-120}
               \hbox{\today}
\vss}}%
}
\vskip20mm
\begin{center}
{\large\bf Perturbation Analysis of Deformed Q-Ball and Primordial
Magnetic Field}
\end{center}

\begin{center}
{\large Tetsuya Shiromizu\footnote
{JSPS Postdoctal Fellowship for Research Abroad}} \\
\vskip 3mm
\sl{DAMTP, University of Cambridge \\ 
Silver Street, Cambridge CB3 9EW, UK \\
\vskip 5mm
Department of Physics, The University of Tokyo, Tokyo 113-0033, Japan \\
and \\
Research Center for the Early Universe (RESCEU), \\ The University of Tokyo, 
Tokyo 113-0033, Japan
}
\vskip 3mm
{\large Tomoko Uesugi \footnote{Research Fellow of the Japan Society 
for the Promotion of Science.}} \\
\vskip 3mm
\sl{Institute for Cosmic Ray Research, 
The University of Tokyo, Tokyo 188-8502, Japan}\\
\vskip 3mm
{\large Mayumi Aoki $^\dagger$} \\
\vskip 3mm
\sl{Graduate School of Humanities and Sciences, 
Ochanomizu University, Tokyo 112-8610, Japan}
\end{center}

\begin{center}
{\it to be published in Phys. Rev. D}
\end{center}

\begin{abstract} 
We study the excited states of the Q-balls by performing stationary 
perturbation on the spherical Q-balls. We find the exact solution 
of the stationary perturbation of the global Q-ball with 
thin wall approximation. 
For local Q-balls we solve the equations of motion for the
perturbative part approximately by using expansion about the coupling 
constant. Furthermore we comment on the magnetic field 
generated by the excited Q-balls during the phase transition 
precipitated by solitosyhthesis and give an implication into cosmology. 
\end{abstract}
\vskip1cm


\section{Introduction}

The existence of the coherent magnetic field over various 
astrophysical scales has been established up to the present\cite{Obs}. 
To explain the observed magnetic field for galaxies, 
clusters and inter-clusters, we need a generation mechanism 
of the seed of primordial magnetic fields\cite{Olinto}. This is because  
magnetohydrodynamics shows that the magnetic fields cannot 
exist unless there is non-zero field strength as a initial 
condition. 

It is known that 
if the seed of magnetic fields with $10^{-19}{\rm Gauss}$ exists 
over the present comoving scale of the proto galaxy, 
$\sim 100{\rm kpc}$, these fields might be amplified to the observed galactic 
magnetic fields, $10^{-6}$Gauss, by the dynamo mechanism \cite{Dynamo}
\footnote{Kulsrud and Anderson pointed out that the kinetic 
dynamo theory breaks down in the interstellar mediums\cite{KA}. 
Whether the mechanism can work sufficiently or not is in the 
debate.}. 

Recently, several generation mechanism of the magnetic field 
in the electroweak phase transition have been proposed. 
One of them is based on the 
thermal fluctuations\cite{Thermal}. However, this mechanism cannot 
provide enough coherent scale of the magnetic field,   
and the net magnetic field will be too small to explain the present 
observation. Another mechanism is founded on the bubble nucleation
\cite{Bubble}. This mechanism is able to supply the large coherent
length of the magnetic field and it can provide the enough field
strength for the onset condition of the dynamo mechanism. 
In this case, however, one needs to assume that the phase transition 
is strongly first order one and completed by the expansion of the 
critical bubbles. Although several attempt to clarify the detail of the phase 
transition have been performed, the order of the phase transition 
has not been clear at this stage \cite{Shiromizu}. 

In this paper, we considered the magnetogenesis due to excited 
states of Q-balls during the phase transition, suggested by one of 
the present authors \cite{Tetsuya}. 
Q-ball is non-topological soliton solution of the complex scalar field 
arising in the theory with an unbroken continuous symmetry
\cite{Coleman} \cite{TDLee}.
The generation of Q-ball\cite{Soliton} and the possibility of the phase 
transition precipitated by solitosynthesis \cite{Kusenko}\cite{Alex}
have been actively investigated. These studies are based on the ground 
state of Q-ball.  If we consider the excited states of Q-ball,
however, it is conceivable that Q-ball has angular momentum. 
Then, it could also have the magnetic moment if it is the 
gauged Q-ball \cite{Lee}. 

We re-analyze the excited state of the Q-ball by 
the stationary perturbation analysis, not by the surface wave
investigated by Coleman\cite{Coleman}. We should note that 
the stationary perturbation is not general. However, it is still 
worth to investigate them because one may expect that the 
thermal distribution is decided by stationary states. We found that 
the $\ell=1$ mode which does not 
exist as in the case of surface wave. 
The result might have some important effects on cosmology, 
since the contribution of the $\ell=1$ mode is larger than the 
model $\ell=2$ in general. Furthermore, taking account of the above 
fact and assuming copious production of gauged Q-balls, 
we estimate the magnetic field of the excited Q-balls. 

The rest of the present paper is organized as follows. In Sec. II, and 
III, we analyze the stationary perturbation on the global Q-balls and local
Q-balls of the ground states and estimate the conserved 
quantities such as the angular momentum and the magnetic field etc. 
In the subsections of the above two sections, we review the feature of 
the Q-balls in ground state, and we also give a sketch of 
the phase transition due to gauged Q-balls. 
Finally, we give an implication into cosmology in Sec. IV. 

\section{Global Q-Balls}


\subsection{Review of Q-Balls in the Ground State}

In this subsection we review briefly global Q-balls in the ground state. 
For simplicity, we consider only the complex scalar field with a global 
U(1) symmetry.   
According to the Coleman's study, the Q-ball has stationary and spherical 
configuration,
%
\begin{equation}
\phi_0 =\varphi_0(r)e^{i \omega t}.
\end{equation}
%
For the existence of the Q-ball solution, we consider here the 
potential which has two minima, that is to say, true vacuum and false 
vacuum.  
In the thin wall approximation, the spatial configuration of the 
scalar field, $\varphi_{0}(r)$, is characterized by the radius of 
Q-ball, $R$, the wall width, $\delta$, and the field values at the 
center of configuration, $\sigma_+$. Here we set the field value at the 
minima: $0$, $\sigma_+$, and the top of the potential barrier: 
$\sigma_-$.  
Then the total energy can be written by 
%
\begin{eqnarray}
E_0 & = & 
\frac{Q_0^2}{2\int \varphi^2_0d^3x}+\frac{1}{2}\int (\nabla \varphi_0)^2d^3x
+\int U(\varphi_0) d^3x \nonumber \\
& = & \frac{3}{8\pi} \frac{Q_0^2}{R^3\sigma_+^2}+2\pi R^2\delta \Bigl( 
\frac{\sigma_+}{\delta}\Bigr)^2+4\pi R^2 \delta 
U_- +\frac{4\pi}{3}U_+R^3,
\end{eqnarray}
%
where $U_\pm:=U(\sigma_\pm)$ and $Q_0$ is the total conserved charge:  
$Q_0=\int d^3 x \psi_0^2$. 

The most favourable configuration is determined by the variational 
principle to minimize the total energy. From 
$\partial E_0/\partial \delta|_{\delta=\delta_*}=0$, the wall width has the 
expression 
%
\begin{equation}
\delta_*=\frac{1}{{\sqrt {2}}} \frac{\sigma_+}{{\sqrt {U_-}}}.
\end{equation}
%
Thus, the total energy becomes
%
\begin{equation}
E_{0*}=E_0(\delta_*)=\frac{3}{8\pi}\frac{Q_0^2}{R^3\sigma_+^2}
+4 {\sqrt {2}}\pi \sigma_+{\sqrt {U_-}}R^2
+\frac{4\pi}{3}U_+R^3.
\end{equation}
%
If the potential has two degenerate minima, $U_+=U(0)=0$, 
the total energy becomes
%
\begin{equation}
E_0=\frac{3}{8\pi}\frac{Q_0^2}{R^3\sigma_+^2}+4 {\sqrt {2}}\pi 
\sigma_+{\sqrt {U_-}}R^2.
\end{equation}
%
 From $\partial E_0/ \partial R|_{R=R_*}=0$, one obtains the radius 
at which the energy takes minimum value,
%
\begin{equation}
R_0^*=\Bigl(\frac{9}{64{\sqrt {2}}\pi^2}\Bigr)^{1/5}
\frac{Q_0^{2/5}}{\sigma_+^{3/5}U_-^{1/10}}.
\end{equation}
%
Then, the total energy becomes
%
\begin{equation}
E_0^*=E_{0*}(R_0^*)=
\frac{15}{16 \pi} \Bigl( \frac{64{\sqrt {2}}\pi^2 }{9} \Bigr)^{3/5}
\frac{Q_0^{4/5}U_-^{3/10}}{\sigma_+^{1/5}}.
\end{equation}
%
For simplicity, we write $R_0^*$ as $R$ hereafter.  



\subsection{Stationary Perturbation on Q-Balls}

Let us consider the stationary perturbation on the Q-balls. The 
perturbation was analyzed in Ref. \cite{TDLee} from the 
viewpoint of the stability. In this paper, we give explicit solutions 
of the stationary perturbation. Although we will not give any implication 
to cosmology in this calculation on the global Q-balls, the analysis 
give a educational examination for the extension into the local 
Q-ball in which we can give an implication into cosmology. 
    
The Lagrangian of the complex scalar fields is 
%
\begin{equation}
L=\int d^3 x \Bigl[ \frac{1}{2}|{\dot \phi}|-\frac{1}{2}|\nabla \phi|^2
-U(|\phi|) \Bigr],
\end{equation}
%
where $U(|\phi|)$ is the potential. The energy momentum tensor becomes
%
\begin{equation}
T_{\mu \nu} = \frac{1}{2}\nabla_\mu \phi^* \nabla_\nu \phi
+\frac{1}{2}\nabla_\mu \phi \nabla_\nu \phi^*-g_{\mu \nu}
\Bigl[\frac{1}{2}\nabla^\rho \phi^* \nabla_\rho \phi +U(|\phi|) \Bigr]
\end{equation}
%

We add the stationary perturbation on the Q-ball in the ground state 
as follows,
%
\begin{equation}
\phi = \phi_0+\phi_1= 
\Bigl( \varphi_0(r)+ {\sqrt {4\pi}} 
\varphi_{\ell m}(r)Y_{\ell m}(\theta, \varphi) 
\Bigr)e^{i\omega t},
\end{equation}
%
where $\ell \geq 1$ and $\ell \geq |m|$. Then the Lagrangian becomes
%
\begin{eqnarray}
L & = & L_0+L_1 \nonumber \\
  & = & \int d^3 x \Bigl[ \frac{1}{2}\omega^2\varphi_0^2
-\frac{1}{2}(\nabla \varphi_0)^2-U(\varphi_0) \Bigr] 
+\frac{1}{2}\int d^3 x \Bigl[ - \Bigl(\frac{d \varphi_{\ell m}}{dr}
\Bigr)^2-V_\ell(r) (\varphi_{\ell m})^2 \Bigr],
\end{eqnarray}
%
where 
%
\begin{equation}
V_\ell(r)=U''(\varphi_0)-\omega^2+\frac{\ell ( \ell +1)}{r^2}.
\end{equation}
%
One can see easily from the eq. (2.11) 
that the perturbation $\varphi_{\ell m}$ satisfies
%
\begin{equation}
-\frac{d^2\varphi_{\ell m}}{dr^2}-\frac{2}{r}\frac{d\varphi_{\ell m}}{dr}
+V_\ell(r) \varphi_{\ell m}=0.
\end{equation}
%
The above equation is similar to the Schr\"{o}dinger equation with 
the potential $V_\ell(r)$ in the ground state. 
In the thin wall approximation, one can set  
%
\begin{equation}
U''(\psi_0(r))=\mu_0^2 \theta(r-R)+\mu^2 \theta (R-r),
\end{equation}
%
where $\mu_0^2 = U''(0)$ and $\mu^2 = U''(\sigma_+)$, and then the 
eq. (2.13) can be solved exactly.  For $r<R$, the
solution is  
%
\begin{eqnarray}
\varphi_{\ell m}(r) & = & \varphi_\ell^{<}(r)=
A'_\ell j_\ell (x) 
\nonumber \\
& = & A_\ell r^\ell \Bigl(\frac{1}{r}\frac{d}{dr} \Bigr)^\ell 
\Bigl( \frac{{\rm sin}(\lambda r)}{r}  \Bigr),
\end{eqnarray}
%
where $x= \lambda r$ and $\lambda ={\sqrt {\omega^2 - \mu^2}}$.  For $r>R$, 
%
\begin{eqnarray}
\varphi_{\ell m}(r)& = & \varphi_\ell^{>}(r)
= B'_\ell h^{(1)}_\ell (y) \nonumber \\
& = & B_\ell r^\ell \frac{d^\ell}{d(r^2)^\ell}\Bigl( \frac{e^{-\gamma r}}
{r}\Bigr)
\end{eqnarray}
%
where $y = i \gamma r$ and $\gamma = {\sqrt {\mu_0^2 -\omega^2}}$. Here 
remember that the condition $\mu_0 > \omega > \mu$ is necessary for 
the existence of the stable Q-ball\cite{Coleman}\cite{TDLee}.  

First, we consider the $\ell=1$ case. In this case the solution becomes 
%
\begin{eqnarray}
 \varphi^{<}_1(r) =\varphi_{1 m}(r)= \frac{A_1}{r^2} 
\Bigl( - {\rm sin}(\lambda r)
+ \lambda r {\rm cos}(\lambda r) \Bigr)
\end{eqnarray}
%
and 
%
\begin{eqnarray}
\varphi_{1 m}(r)= \varphi^{>}_1(r) =\frac{B_1}{2r^2} 
\Bigl( - \gamma r -1 \Bigr)e^{-\gamma r}. 
\end{eqnarray}
%
The relation between the prefactors, $A_1$ and $B_1$, is determined by 
the matching conditions of the field on the wall. 
As the potential $V_\ell$ is continuous, one should require the 
$\varphi^{>}_1(R)=\varphi^{<}_1(R)$ 
and  $\varphi'^{>}_1(R)=\varphi'^{<}_1(R)$. Each requirements imply  
%
\begin{eqnarray}
(1+\gamma R)e^{-\gamma R}B_1=
2\Bigl[ {\rm sin}(\lambda R)-\lambda R {\rm cos}(\lambda R )  \Bigr]A_1,
\end{eqnarray}
%
and 
%
\begin{eqnarray}
A_1 \Bigl( -2 \lambda R {\rm cos}(\lambda R )+2 {\rm sin}(\lambda R)
-(\lambda R)^2 {\rm sin}(\lambda R) \Bigr)
=B_1\Bigl( \frac{1}{2}(\gamma R)^2+\gamma R +1 \Bigr)e^{-\gamma R}.
\end{eqnarray}
%
Thus, one finds 
%
\begin{eqnarray}
\frac{-2\alpha {\rm cos} \alpha +(2-\alpha^2){\rm sin} \alpha}
{{\rm sin}\alpha-\alpha {\rm cos}\alpha}
=\frac{\beta^2+2\beta +2}{1+\beta},
\end{eqnarray}
%
where $\alpha=\lambda R$ and $\beta = \gamma R $. 
For $R \mu =1$ and $R\mu_0 =4$, the eigenvalue $\omega_1$ is 
$\omega_1 \simeq 3.6 R^{-1}$. For $R \mu =1$ and $R\mu_0 =5$, 
the eigenvalue $\omega_1$ is $\omega_1 \simeq 3.8 R^{-1}$. The perturbation 
is permitted only in the case of the discretized frequency. Note that 
$\ell=1$ modes of the surface wave do not exist in the Coleman's
study. As $\ell=1$ modes can be easily created than $\ell=2$ modes in
general, the existence of the $\ell=1$ modes is important in the context
of cosmology. 

Next, we give some conserved quantities which specify the feature 
of the perturbation. The charge of the perturbation is 
%
\begin{eqnarray}
\delta Q_m & = & \int d^3x j_0 = \omega_1 \int d^3x \varphi_1^2
\nonumber \\
& = & A_1^2 \frac{\omega_1}{R}\Bigl[ -\frac{1}{2}+\frac{1}{2}\alpha^2
+\frac{1}{2}{\rm cos}(2 \alpha) +\alpha \frac{1}{4}{\rm sin}(2 \alpha)+
\Bigl( \frac{{\rm sin}(\alpha)-\alpha {\rm cos}(\alpha)}
{1+\beta}  \Bigr)^2 \Bigl( 1+\frac{1}{2}\beta \Bigr) \Bigr] .
\end{eqnarray}
%
The angular momentum $J_{1m}$ is 
%
\begin{eqnarray}
J_{1m}=\int d^3 x T_{0 \varphi} = m \omega_1 \int d^3 x \varphi_1^2=m 
\delta Q.
\end{eqnarray}
%

Finally, we consider the $\ell =2 $ case. From the eqs. (2.15) and 
(2.16) the solutions are 
%
\begin{eqnarray}
\varphi_2^{<}(r) = \varphi_{2 m}(r) = \frac{A_2}{r^3}\Bigl[   \
3{\rm sin}(\lambda r)-3 \lambda r {\rm cos}(\lambda r) 
-(\lambda r)^2 {\rm sin}(\lambda r) \Bigr]
\end{eqnarray}
%
and
%
\begin{eqnarray}
\varphi_2^{>}(r) = \varphi^{2 m}(r)= \frac{B_2}{4r^3}
\Bigl( 3+3 \gamma r +\gamma^2 r^2   \Bigr) e^{-\gamma r}.
\end{eqnarray}
%

The matching conditions, $\varphi_2^{<}(R) =\varphi_2^{>}(R)$ and 
${\varphi'}_2^{<}(R) = {\varphi'}_2^{>}(R)$ at $r=R$, give the relation 
%
\begin{eqnarray}
\frac{(3-\alpha^2){\rm sin}\alpha -3 \alpha {\rm cos}\alpha}
{(-9+4 \alpha^2){\rm sin}\alpha +\alpha (9-\alpha^2){\rm cos}\alpha}
=-\frac{3+3\beta+\beta^2}{9+9\beta+4\beta^2+\beta^3}
\end{eqnarray}
%
For $\mu R =1$ and $\mu_0 R=4$, the solution of $\omega$ does not
exist. This comes from the fact that the bottom of the potential
$V_\ell$ raises as $\ell$ increase and then decrease the number 
of the bound state.  
For $\mu R =1$ and $\mu_0 R=5$, one has the solution 
$\omega_2 \simeq 4.66R^{-1}$.


\section{Excited state of Local Q-Balls}

So far, we have analyzed the perturbation on the global Q-balls. 
In this section, we perform the perturbation analysis on the local
Q-balls coupled to a U(1) gauge field.  This analysis will be  
significant to cosmology because such object might be able to supply 
a seed of astrophysical magnetic field.

\subsection{Local Q-balls in the Ground State and Phase Transition}

The basic features of the local Q-ball 
was studied by Lee et al in Ref.\cite{Lee}. 
We consider the theory with a complex scalar field $\phi$ coupled to 
a U(1) gauge field $A_{\mu}$.   
The Lagrangian density is 
%
\begin{eqnarray}
{\cal L} = \frac{1}{2}|(\partial_\mu-ie A_\mu)\phi|^2-U(|\phi|)
-\frac{1}{4}F_{\mu\nu}F^{\mu\nu},
\end{eqnarray}
%
where $F_{\mu\nu}=\partial_\mu A_\nu-\partial_\nu A_\mu$.
The conserved current and the energy momentum tensor 
which induce the conserved quantities are
%
\begin{eqnarray}
j_\mu=\frac{i}{2}[\phi(\partial_\mu+ieA_\mu)\phi^*-
\phi^*(\partial_\mu-ieA_\mu)\phi ]
\end{eqnarray}
%
and
%
\begin{eqnarray}
T_{\mu\nu}=\frac{1}{2}\Bigl[ (\partial_\mu-ieA_\mu)\phi 
 (\partial_\nu+ieA_\nu)\phi^* + (\partial_\nu-ieA_\nu)\phi 
(\partial_\mu+ieA_\mu)\phi^*\Bigr] - F_{\mu
 \sigma}F_\nu^\sigma-{\cal L}\eta_{\mu\nu},
\end{eqnarray}
%
respectively. 

To find the Q-balls in the ground state, we seek for the solution such that 
%
\begin{eqnarray}
\phi=\varphi(r)e^{i\omega t}~~~~~~~{\rm and}~~~~~~~ A_\mu=A_0(r)\delta_{\mu 0}
\end{eqnarray}
%
Using the above configurations, the Lagrangian becomes 
%
\begin{eqnarray}
L=4 \pi \int dr r^2 \Bigl[ -\frac{1}{2}\varphi'^2+\frac{1}{2e^2}
g'^2+\frac{1}{2}g^2 \varphi^2-U(\varphi)\Bigr], 
\end{eqnarray}
%
where $g=\omega - e A_0(r)$. The conserved charge associated with 
the U(1) symmetry becomes 
%
\begin{eqnarray}
Q=\int d^3x j_0= \int d^3 x (\omega - e A_0)\varphi^2
=\int d^3x g \varphi^2 .
\end{eqnarray}
%
Assumed the thin wall approximation, as well as the case of global 
Q-balls, the configuration 
of the scalar field, $\varphi$, is roughly expressed by the radius,
$R$, and the field value at the minimum of the potential, $\sigma_+$. 
\footnote{Rigourously speaking, the value of the scalar field $\varphi$ 
for $r < R$ should be determined by
the variational principle with respect to the expression for 
the total energy $E$ \cite{Lee}. 
One can approximate it as $\sigma_+$, however, if only the 
leading-order term is considered.
The radius $R$ is decided so that the energy of the 
non-trivial configuration takes minimum value. }. 

It is given by      
%
\begin{eqnarray}
\varphi(r)=\sigma_+ \theta (R-r) .
\end{eqnarray}
%
In this case, we can solve the equation of motion for $g$ \cite{Lee}.  
The solution becomes   
%
\begin{eqnarray}
& &g(r)=\frac{\Bigl( \omega - \frac{e^2 Q}{4\pi R} \Bigr)R}{{\rm sinh}
(e\sigma_+ R)} \frac{{\rm sinh}(e \sigma_+ r)}{r} 
~~{\rm for}~~~r \leq R, \\ 
& &g(r)=\omega - \frac{e^2Q}{4\pi r} \ \ \ \ \ \ ~~~~~{\rm for}~~~~r > R .
\end{eqnarray}
%
Inserted the above solution 
into the eq. (3.6), the relation between $Q$ and $\omega$ is determined
by  
%
\begin{eqnarray}
\omega = \frac{e^2 Q}{4\pi R}\frac{1}{1-\frac{{\rm tanh}x}{x}},
\end{eqnarray}
%
where $x=e\sigma_+ R$. 
The total energy of the Q-ball becomes\footnote{Although the
configuration of the scalar field is approximately
expressed by the step function, we considered the surface term 
with the finite width here. Readers might feel a confusion. However, 
after all, if we assume $e\sigma_+ R >1$, one see that the 
inconsistency of the approximation can vanish.}
%
\begin{eqnarray}
E& =& \frac{1}{2}\omega Q+4\pi \int dr r^2 \Bigl[
\frac{1}{2}\varphi'^2+U(\varphi)
\Bigr]\nonumber \\
& \simeq & \frac{e^2 Q^2}{8\pi R}\Bigl( 1-\frac{{\rm tanh}x}{x}\Bigr)^{-1}
+4{\sqrt {2}}\pi \sigma_+{\sqrt{U_-}}R^2+  \frac{4\pi}{3}U_+ R^3,
\end{eqnarray}
%
If $e\sigma_+ R > 1$, 
as will be discussed later,
the energy can be approximately written by 
%
\begin{eqnarray}
E \simeq \frac{e^2Q^2}{8\pi R}+4{\sqrt {2}}\pi \sigma_+{\sqrt
{U_-}}R^2 +\frac{4\pi}{3}U_+ R^3.
\end{eqnarray}
%

Let us consider the phase transition.  We note that our 
study is different from the previous one by Ellis et al \cite{Ellis}. 
In Ref. \cite{Ellis}, they discussed the phase transition which occurs 
above the critical temperature, and took no account 
of the existence of the critical charge. We discuss the 
phase transition which takes place below the critical temperature 
and above the temperature at which the standard nucleation\cite{Linde}
starts. 

As the universe expands, the temperature of the universe cools down. 
At the critical 
temperature $T_c$, the potential has the two degenerate 
minima at $\varphi=\sigma_{+}$ and $\varphi=0\ ; U_+=U(0)=0$. 
In this case, the total energy becomes 
%
\begin{eqnarray}
E=\frac{e^2Q^2}{8\pi R}+4{\sqrt {2}}\pi\sigma_+{\sqrt {U_-}}R^2.
\end{eqnarray}
%
There exists the upper limit on charge and radius of Q-ball 
\cite{Lee}. 
 From $\partial E /\partial R|_{R=R_*}=0$, the radius of Q-ball is
given by   
%
\begin{eqnarray}
R_*=\Bigl(  \frac{e^2Q^2}{64{\sqrt {2}}\pi^2\sigma_+{\sqrt {U_-}}}
\Bigr)^{1/3}.
\end{eqnarray}
%
Inserted this equation into the eq. (3.13), the minimum energy 
becomes
%
\begin{eqnarray}
E_*=E(R_*)=\frac{3}{2\cdot 2^{5/6}\pi^{1/3}}(\sigma_+{\sqrt
{U_-}})^{1/3}(eQ)^{4/3}.
\end{eqnarray}
%
Here note that $E_*$ is the monotonically increasing function 
of $Q$. To construct stable Q-balls, however, the total energy $E_*$ cannot 
exceed the energy for $Q$'s free particles with the mass 
$\mu$, 
otherwise the Q-balls will decay into $Q$'s free particles. 
Thus, from the condition 
$\mu= \partial E_* /\partial Q|_{Q=Q_{\rm max}}$, one can obtain the 
upper bound on the total charge $Q_{\rm max}$ as
%
\begin{eqnarray}
Q_{\rm max}=\frac{\pi}{{\sqrt {2}}e^4}\frac{\mu^3}{\sigma_+{\sqrt {U_-}}}.
\end{eqnarray}
%
The upper bound on the radius $R_{\rm max}$ of Q-ball is also given by
%
\begin{eqnarray}
R_{\rm max}=\frac{1}{4{\sqrt {2}}e^2}\frac{\mu^2}{\sigma_+{\sqrt
{U_-}}}. 
\end{eqnarray}
%

We now consider the temperature $T_q (< T_c)$. 
We assumed that $T_q$ is higher than the temperature at which the 
bubble nucleation starts. 
In Ref. \cite{Alex}, it was pointed out that the Q-ball has the 
critical charge $Q_c$. If the charge exceeds this critical value, 
Q-ball can expand. 
If the charge is smaller than the critical one, 
there exists the range of the total energy in which the 
Q-ball is bounded. 
At $T=T_q$, if Q-ball is able to have the critical charge,   
it is possible that the phase transition is 
precipitated by the Q-ball growing up into the macroscopic size. 
For the Q-ball with critical charge, 
various quantities are decided by the variational principle
so as to satisfy the conditions 

%
\begin{equation}
\partial E/ \partial R|_{R=R_c,Q=Q_c}=\partial^2E/\partial R^2|_{R=R_c,Q=Q_c}= 0 .
\end{equation}
%
Resultant expressions for $Q_c, R_c,$ and $E_c$ are respectively
obtained as
%
\begin{eqnarray}
Q_c=6{\sqrt {6}}\pi\frac{\sigma_+^2U_-}{e|U_+|^{3/2}},
\end{eqnarray}
%
%
\begin{eqnarray}
R_c=\frac{3}{{\sqrt {2}}}\frac{\sigma_+{\sqrt {U_-}}}{|U_+|},
\end{eqnarray}
%
and
%
\begin{eqnarray}
E_c=18{\sqrt {2}}\pi \frac{\sigma_+^3U_-^{3/2}}{|U_+|^2}.
\end{eqnarray}
%
As in the case of $T=T_c$, the energy must satisfy 
%
\begin{eqnarray}
E_c/Q_c={\sqrt {3}}e\sigma_+\Bigl( \frac{U_-}{|U_+|}\Bigr)^{1/2} < \mu.
\end{eqnarray}
%
According to Kusenko\cite{Alex}, the large charge can be attained soon
and then the almost of Q-balls has the maximum 
charge $Q_{\rm max}$. If $Q_{\rm max} > Q_c$, Q-ball can 
expand as soon as the temperature becomes lower than the critical one 
because the growth rate of the charge is $\sim \eta_\varphi n_\gamma 
(4\pi R_c^2){\sqrt {T/m_\varphi}} \sim 10^{-7} {\rm GeV} 
\gg H \sim 10^{-15}{\rm GeV}$ for TeV scale. 
Thus, the expansion of the Q-ball will start before the nucleation of 
the critical bubble. To compare $Q_c$ with $Q_{\rm max}$ ,
let us consider the ratio
%
\begin{eqnarray}
\frac{Q_c}{Q_{\rm max}} = 12{\sqrt {3}}e^3 \Bigl( \frac{\sigma_+ {\sqrt
{U_-}}}{\mu |U_+|^{1/2}}\Bigr)^3, 
\end{eqnarray}
%
where we used the approximation, $U_-(T_q) \sim U_-(T_c)$.
The condition $Q_c/Q_{\rm max}<1$ implies 
%
\begin{eqnarray}
2^{2/3}{\sqrt {3}}e \sigma_+\Bigl( \frac{U_-}{|U_+|}\Bigr)^{1/2} < \mu.
\end{eqnarray}
%
Comparing eq. (3.22) with eq. (3.24), one can see that the most strict
constraint is the latter one (eq. (3.24)). 

Here we must remember that $e\sigma_+ R>1$ was assumed in the
derivation of the eq. (3.12). 
The conditions
$e\sigma_+ R_{\rm max}>1$ and $e \sigma_+ R_c >1$, therefore,
must be satisfied, so that the coupling constant $e$ is given by
%
\begin{eqnarray}
\frac{{\sqrt {2}}}{3}\frac{|U_+|}{\sigma_+^2{\sqrt {U_-}}}
< e < \frac{1}{4{\sqrt {2}}}\frac{\mu^2}{{\sqrt {U_-}}}.
\end{eqnarray}
%

To see the details, we consider a potential 
%
\begin{eqnarray}
U(\varphi)=\frac{1}{2}\mu^2 \varphi^2-\frac{1}{6}M\varphi^3+
\frac{\lambda}{24}\varphi^4,
\end{eqnarray}
%
where $\mu$, $M$, and $\lambda$ depend on the temperature of the 
universe. 
At the near critical temperature, one can take $y=\lambda^{1/2}(\mu/M)=
y_c+\epsilon$, where $\epsilon$ is a small non-dimensional quantity, 
and $y_c=1/{\sqrt{3}}$ is the value of $y$ at the critical temperature, 
decided by $U_+=0$. 
The leading order of these quantities become
%
$ U_+ \simeq \frac{\sigma_+^2M^2 y_c}{\lambda} \epsilon, 
%
%
U_- \simeq \frac{1}{24}\frac{\sigma_-^2M^2}{\lambda},
%
%
\sigma_+ \simeq \frac{2M}{\lambda},
%
%
\sigma_- \simeq \frac{M}{\lambda},
%
~{\rm and}~~
%
\mu \simeq y_c\frac{M}{{\sqrt {\lambda}}}$.
%
Substituting the above equations into eq. (3.24), 
one can obtain the constraint on 
the coupling constant $e$; 
%
\begin{eqnarray}
e< \frac{2^{5/6}}{3^{3/4}} 
 {\sqrt {\lambda |\epsilon |}}.
\end{eqnarray}
%
Furthermore, from the eq. (3.24), the range where the present approximation
is valid becomes
%
\begin{eqnarray}
\frac{4}{3} \lambda^{1/2} |\epsilon |< e < 
\frac{1}{2{\sqrt {3}}}\lambda^{1/2} .
\end{eqnarray}
%
We can see that 
the constraint (eq. (3.27)) can work for a sufficiently small value of
$\epsilon $. 

We investigated the phase transition caused by 
not global Q-ball but local Q-ball. 
To accomplish the phase transition caused by local Q-ball,
we showed that the coupling constant $e$ must be too small. In other
words, if the model satisfies 
%
\begin{eqnarray}
\frac{3^{3/2} e^2}{2^{5/3}}< \lambda |\epsilon |,
\end{eqnarray}
%
the charge of Q-ball can exceed the critical value as soon as  
the universe cools down under the critical temperature. 
In this situation, the phase transition will go as follows. 
By solitosynthesis, Q-balls with maximum charge $Q_{\rm max}$ 
is created above the critical temperature. 
If the coupling constant satisfies the above constraint (eq.(3.29)), 
the charge of Q-ball exceeds the critical value, $Q_c$.    
Then all the Q-balls must expand to macroscopic size at the 
temperature lower than the critical one. 
Such phenomena always happen before bubble nucleation starts. 
Thus, the new picture of the phase transition 
due to gauged Q-balls could become real one.   

We note that the present scenario of the phase transition 
is slightly different from the Kusenko's one. In Ref.\cite{Alex}, 
the accretion of charge which happens only below the critical 
temperature was considered. However, it is conceivable that such 
mechanism will take place above the critical temperature too, and 
we have taken account of it.

\subsection{Stationary Perturbation on Local Q-balls}

Let us consider the stationary perturbation on the local Q-balls. 
The perturbed quantities are written by 
%
\begin{eqnarray}
\phi=(\varphi_0+{\sqrt {4\pi}}\varphi_{\ell m} Y_{ \ell m})e^{i \omega t}
\end{eqnarray}
%
and
%
\begin{eqnarray}
A_\mu = {}^{(0)}A_\mu+{}^{(1)}A_\mu
\end{eqnarray}
%
We assume the following ansatz on perturbation of the vector potential: 
${}^{(1)}A_0={}^{(1)}A_r =0$, and 
$ {}^{(1)}A_I = \frac{{\sqrt {4\pi}}}{i}g_{\ell m}(r)\partial_I(Y_{\ell m}-
Y^*_{\ell m})$\footnote{Note that all other types of the expression
cannot couple with $\varphi_{\ell ,m}$.}, where suffix $I$ runs over $\theta , \varphi$. 

The Lagrangian for the 
perturbations becomes 
%
\begin{eqnarray}
\delta L & = & \int d^3x \Bigl[ -\frac{1}{2}(\varphi'_{\ell m})^2
+\frac{1}{2}g_0^2 \varphi_{\ell m}^2-\frac{1}{2}
U''(\varphi_0) \varphi_{\ell m}^2-\frac{\ell(\ell
+1)}{2r^2} \varphi_{\ell m}^2 \Bigr] \nonumber \\
& & +\int d^3 x \Bigl[ -(g'_{\ell m})^2 -e^2\varphi_0^2 g_{\ell m}^2
+e g_{\ell m}\varphi_0 \varphi_{\ell m} \Bigr]\frac{\ell (\ell +1)}{r^2}.
\end{eqnarray}
%
Thus, the equations of the perturbation become
%
\begin{eqnarray}
-g_{\ell m}''+e^2\varphi_0^2g_{\ell m}=\frac{1}{2}e \varphi_0 
\varphi_{\ell m}
\end{eqnarray}
%
and
%
\begin{eqnarray}
\Bigl[ -\frac{d^2}{dr^2}-\frac{2}{r}\frac{d}{dr}
-g_0^2+U''(\varphi_0)+\frac{\ell (\ell +1)}{r^2}\Bigr] 
\varphi_{\ell m}=\frac{e \ell 
(\ell+1)}{r^2} \varphi_0 g_{\ell m},
\end{eqnarray}
%
respectively. 

The total charge, the angular momentum and the total energy are 
given by 
%
\begin{eqnarray}
Q=\int d^3x j_0=\int d^3x g_0(\varphi_0^2+\varphi_{\ell m}^2)=Q_0 +\delta Q_m,
\end{eqnarray}
%
%
\begin{eqnarray}
J_m= \int d^3x T_{0 \varphi}
=m\int d^3x g_0 \Bigl[ \varphi_{\ell m}^2- e g_0 \varphi_0 g_{\ell m} 
\varphi_{\ell m} \Bigr],
\end{eqnarray}
%
and
%
\begin{eqnarray}
E& = & \frac{1}{2}\omega Q_0+\int d^3x 
\Bigl[ \frac{1}{2}\varphi_0'^2+U(\varphi_0) \Bigr] 
+\frac{1}{2}\omega \delta Q_m +\frac{1}{2}\int d^3 x
g_0^2\varphi^2_{\ell m} \nonumber \\
& =:& E_0+\delta E,
\end{eqnarray}
%
where $\delta E =(1/2)\omega \delta Q_m +(1/2)
\int d^3xg_0^2\varphi^2_{\ell m}$. 

As we stated in Introduction, the perturbed Q-ball has the magnetic field. 
The averaged magnetic field $B^i= \epsilon^{ijk}\partial_j A_k$ 
over the angular direction is 
%
\begin{eqnarray}
{\overline {B^2}}(r):=\frac{1}{4\pi} \int d\Omega {\bf B}^2 
= 2\ell (\ell +1) \frac{{g'}_{\ell m}^2}{r^2}. 
\end{eqnarray}
%
We consider only the $\ell=1$ mode of excited Q-balls, since it is  
unrealistic that the Q-balls with much higher excitations actually 
affect. For the order estimation, we assume $e \sigma_0 R \ll 1$. In this 
case $g^{<}_0 \sim \omega -e^2Q_0/4\pi R$. Furthermore, we also 
suppose the small coupling constant, $e \ll 1$, and expand the 
part of perturbation in powers of $e$,  
%
\begin{eqnarray}
\varphi_1=\varphi_{1 m}={}^{(0)}\varphi_1+e{}^{(1)}\varphi_1~~~~~~
{\rm and}~~~~~~g_1=g_{\ell m}={}^{(0)}g_1+e{}^{(1)}g_1.
\end{eqnarray}
%
The equations in the order $O(e^0)$ become
%
\begin{eqnarray}
\Bigl[-\frac{d^2}{dr^2}-\frac{2}{r}\frac{d}{dr}+U''(\varphi_0)-\omega^2
+\frac{2}{r^2} \Bigr]{}^{(0)}\varphi_1 =0~~~~{\rm and}~~~~
\frac{d^2}{dr^2} {}^{(0)}g_1=0.
\end{eqnarray}
%
The solutions of these equations become 
%
\begin{eqnarray}
{}^{(0)}\varphi_1^{<}=A_1\frac{d}{dr}\frac{{\rm sin}(\lambda r)}{r}~~~~
{\rm and}~~~~ {}^{(0)}\varphi_1^{>}=B_1r\frac{d}{dr^2}\frac{e^{-\gamma r}}{r}
\end{eqnarray}
%
and
%
\begin{eqnarray}
{}^{(0)}g_1=0
\end{eqnarray}
%
Assumed that $\alpha \ll 1$ and $\beta \ll 1$, the expression 
of $\delta Q_1$ becomes
%
\begin{eqnarray}
\delta Q_1= \frac{2}{15}
\pi A_1^2 \frac{\omega}{R} \alpha^6.
\end{eqnarray}
%
Up to the order of $O(e)$, one can obtain the relation
%
\begin{eqnarray}
{}^{(1)}g_1 \simeq \varphi_0 R^2 {}^{(0)}\varphi_1.
\end{eqnarray}
%
The energy induced by the magnetic field becomes
%
\begin{eqnarray}
E_M & \simeq & 4\pi \int_0^R dr r^2 {\overline {B^2}}(r) \nonumber \\
    & \simeq & \frac{16 \pi}{3}e^2\sigma_+R^3 ({}^{(0)}\varphi_1(R))^2
\nonumber \\
& \simeq & \frac{16\pi}{27}e^2A_1^2\sigma_+^2R^5 \lambda^6 \nonumber
    \\
& \simeq & \frac{40 \pi}{9}e^2Q_0\frac{\sigma_+^2}{\omega},
\end{eqnarray}
%
where we used  $\delta Q \sim Q_0$. 
The mean magnetic field becomes 
%
\begin{eqnarray}
B \simeq {\sqrt {2E_M/\frac{4\pi}{3}R^3 }} \simeq e Q_0^{1/2}
\Bigl( \frac{\sigma_+^2}{\omega R^3}\Bigr)^{1/2},
\end{eqnarray}
%
where $\rho = \omega R$. For the potential 
(3.26), the order of the magnitude becomes 
%
\begin{eqnarray}
B \simeq 24(2\pi)^{1/2}e \frac{\mu^2}{\lambda}\rho^{-1/2}.
\end{eqnarray}
%


\section{An Implication into Cosmology -- Primordial Magnetic Field --}

In this paper, we investigated the perturbation on the ground state 
of Q-balls. 
Under the thin-wall approximation, we solved the equation 
for the part of perturbation analytically in the global Q-balls. 
For the local Q-balls, we solved them by expanding in powers of 
coupling constant $e$. Furthermore, we estimated the mean magnetic 
field for each Q-balls. 
In the context of cosmology, we need to average it. We assume that 
gauged Q-balls can be copiously produced. 
Since it is natural to suppose that the excited states of the Q-ball 
obeys the thermal distribution in such situation\cite{Tetsuya}, 
the cosmological mean magnetic field 
generated by the excitation of the local Q-ball can be evaluated 
by multiplying the factor $e^{-(1/2)\beta \delta E}$, 
where $\delta E$ is the energy of the excitation and is given by 
%
\begin{eqnarray}
\delta E \sim 2 \times \frac{1}{2}\omega \delta Q_1 
\sim \omega Q_0 \sim \frac{4\pi \rho}{e^2}\mu.
\end{eqnarray}
%
Thus, the mean value becomes 
%
\begin{eqnarray}
\langle B \rangle_{R_{\rm max}} \sim 50 \frac{e}{\lambda \rho^{1/2}}
\mu^2 e^{- {\cal E}},
\end{eqnarray}
%
where ${\cal E}:=(1/2)\beta \delta E = \frac{2\pi
\rho}{e^2}\frac{\mu}{T}$. 

At the endpoint of the phase transition, the magnetic field becomes
%
\begin{eqnarray}
\langle B \rangle_{R_f} & \sim & \Bigl( \frac{R_{\rm max}}{R_f} \Bigr)^2
\langle B \rangle_{R_{\rm max}} \simeq 
0.3 \frac{g_* \lambda}{f_b^2e^3 \rho^{1/2}} \frac{T_f^4}{m_{\rm pl}^2}
e^{-{\cal E}} \nonumber \\
& \sim & 10 \times 
\Bigl(\frac{g_*}{100}\Bigr)\Bigl(\frac{\lambda}{1.0}\Bigr)
\Bigl(\frac{f_b}{10^{-3}}\Bigr)^{-2}\Bigl(\frac{e}{0.1}\Bigr)^{-3}
\Bigl( \frac{\rho}{4.0}\Bigr)^{-1/2}\Bigl( \frac{T_f}{100{\rm GeV}}\Bigr)^4
{\rm exp}\Bigl[-800 \pi\Bigl(\frac{\rho}{4.0}\Bigr)\Bigl(\frac{e}{0.1}\Bigr)^{-2}
\frac{\mu}{T}\Bigr]~~{\rm Gauss}. 
\end{eqnarray}
%
where $R_f$ denotes the size of Q-ball at that time and $f_b
=R_f H_f$. Since the Reynolds number becomes ${\rm Re} \sim 10^{12}$ 
\cite{Bubble} at the endpoint of the phase transition, turbulence for 
the electroweak plasma occurs over the scale of the radius of domain 
and then the energy of the 
magnetic field is equipartitioned with the 
energy of the fluid\footnote{Although the amplification due to the
turbulence is not clear at present, it has been used in the 
generation mechanism of the primordial magnetic field during 
the first order phase transition\cite{Bubble}\cite{Sigl}. }. 
Thus the final field strength becomes
%
\begin{eqnarray}
B(R_f)\sim \rho^{1/2} v \sim g_*^{1/2} v T_f^2 \sim 10^{24}
\Bigl( \frac{T_f}{100{\rm GeV}}\Bigr)^2{\rm Gauss},
\end{eqnarray}
%
where $\rho$ and $v$ are the density and the velocity of the fluid, 
respectively. 

If the coherent scale is comoving, the number of magnetic domains inside 
the galaxy scale is $N \sim 10^{10}(10^{-3}/f_b)(T_f/100{\rm GeV})$. 
Thus, averaging over the 
galaxy scale\cite{EO}, one can obtain the present mean value 
%
\begin{eqnarray}
\langle B_{\rm now} \rangle_{\rm galaxy} & \sim & 
\frac{1}{N^{3/2}}\Bigl(\frac{a_{\rm f}}{a_{\rm now}}  \Bigr)^2 B(R_f) 
\nonumber \\
& \sim & 10^{-21}\Bigl( \frac{f_b}{10^{-3}}\Bigr)^{3/2} 
\Bigl( \frac{T_f}{100{\rm GeV}}\Bigr)^{-3/2}
{\rm Gauss}.
\end{eqnarray}
%
The above value satisfies the onset condition of the dynamo mechanism. 

In the above derivation, we assumed for simplicity that the evolution 
of the magnetic field goes only by the red-shift effect. 
Although the definite scenario of the evolution has not been composed 
yet, some studies related with this problem have been done. 
Finally, we will refer to the studies which deal with the details 
of the evolution of the magnetic field. 
In Ref. \cite{Jedamik}, the damping effect due to photon diffusion 
around the recombination era were studied. 
In Ref. \cite{Pattern}, Olesen et al showed that an inverse cascade 
happens if the power of the initial spectrum is larger than $-3$,
and this means that the coherent scale of the magnetic field will be 
able to extend.

\vskip1cm
\centerline{\bf Acknowledgment}
TS is grateful to Gary Gibbons and DAMTP relativity group for 
their hospitality. This work is supported by JSPS fellow. 


\end{document}